\documentclass[11pt,a4]{article}
\pdfoutput=1 
\usepackage{jheppub} 
\usepackage[utf8]{inputenc}
\usepackage{amsfonts}
\usepackage{amsbsy}
\usepackage{url}
\usepackage{enumerate}
\usepackage[justification=raggedright]{caption}
\usepackage{hhline}
\usepackage{graphicx,color}
\usepackage{amssymb,amsmath}
\usepackage{slashed}
\usepackage{hyperref}
\usepackage{braket}
\usepackage{simplewick}
\usepackage{ascmac}
\usepackage{latexsym}
\usepackage{pifont}          
\usepackage{bm}
\usepackage{comment}
\usepackage[normalem]{ulem}
\usepackage{CJKutf8}
\usepackage{cancel}
\graphicspath{{./fig/}{./}{./figs/}}
\usepackage[dvipsnames]{xcolor}

\begin{document}

\preprint{DESY-25-129}

\title{Baryons as linked vortices in QCD matter with isospin asymmetry
}

\author[a,b]{Yu~Hamada,}
\emailAdd{yu.hamada@desy.de}
\affiliation[a]{Deutsches Elektronen-Synchrotron DESY, Notkestr. 85, 22607 Hamburg, Germany}
\affiliation[b]{Research and Education Center for Natural Sciences, Keio University, 4-1-1 Hiyoshi, Yokohama, Kanagawa 223-8521, Japan}

\author[b,c,d]{Muneto~Nitta}
\emailAdd{nitta@phys-h.keio.ac.jp}
\affiliation[c]{Department of Physics, Keio University, 4-1-1 Hiyoshi, Yokohama, Kanagawa 223-8521, Japan}
\affiliation[d]{International Institute for Sustainability with Knotted Chiral Meta Matter(WPI-SKCM$^2$), Hiroshima
University, 1-3-2 Kagamiyama, Higashi-Hiroshima, Hiroshima 739-8511, Japan}

\author[b]{and Zebin~Qiu}
\emailAdd{qiuzebin@keio.ac.jp}

\abstract{
We investigate a baryonic structure in low-energy QCD via a model-independent way using the chiral perturbation theory at the leading order, in the presence of the baryon chemical potential $\mu_B$, the isospin chemical potential $\mu_I$, and the electromagnetic coupling. For such a scenario in the chiral limit, it has been known that the neutral pion winds like in the chiral soliton lattice, confined within an Abrikosov-Nielsen-Olesen (ANO) vortex of the charged pions. This structure undergoes a drastic transformation when the pion mass is introduced, i.e., both charged and neutral pions condense in the bulk, allowing two distinct types of vortices: the charged pions constitute a local ANO-like vortex, while the neutral pion configures a global vortex which is further attached to a domain wall also known as the chiral soliton. Remarkably, the ANO vortex forms a topological linking with the closed global vortex line, when $\mu_B$ exceeds its critical value as a function of $\mu_I$. The linking number has the physical meaning of the baryon number in view of the Wess-Zumino-Witten term. In this sense, the linked configuration realizes a stable Skyrmion-type solution, but innovatively without the Skyrme term. We therefore propose a novel phase of dense baryonic matter comprised of such vortices, which shall play a role in the low-energy QCD phase diagram.
}

\maketitle

\section{Introduction}
The phase structure of Quantum Chromodynamics (QCD) at finite density remains a central but unresolved issue in nuclear physics~\cite{Eichmann:2016yit, Schmidt:2017bjt, Fischer:2018sdj, Rothkopf:2019ipj, Guenther:2020jwe, Braguta:2023yhd, Adam:2023cee}, largely due to the challenge of dealing with strongly correlated systems and the sign problem in Lattice QCD (LQCD). At low energies, effective field theories~\cite{Holt:2014hma,Hammer:2019poc,Drischler:2021kxf} such as chiral perturbation theory (ChPT)~\cite{Gasser:1983yg,Scherer:2002tk} provide useful tools. 
ChPT is constructed in terms of pions as Nambu–Goldstone bosons based on the chiral symmetry breaking.
Topological aspects of ChPT are incorporated in the conserved Goldstone–Wilczek (GW) current~\cite{Goldstone:1981kk}, which couples the baryon current to gauge fields via the Wess–Zumino–Witten (WZW) term~\cite{Witten:1983tw}.
The density effects in the hadronic sector is captured by the baryon chemical potential $\mu_B$ as the temporal component of a fictitious $U(1)_B$ baryon gauge field, 
encoded in the WZW term.
Another relevant gauge field is the electromagnetic $U(1)_{\rm EM}$ gauge field.
Especially, in the presence of a magnetic field, the QCD phase diagram 
exhibits richer structure~\cite{Kharzeev:2015kna,Miransky:2015ava,Andersen:2014xxa,Yamamoto:2021oys,Cao:2021rwx,Iwasaki:2021nrz}. For example, the magnetic catalysis enhances quark condensates at zero temperature~\cite{Klimenko:1990rh}, while inverse magnetic catalysis occurs near the crossover temperature~\cite{Bruckmann:2013oba,Aoki:2006we}. 
Moreover, there are novel phases induced by magnetic fields, such as 
the chiral magnetic spiral~\cite{Basar:2010zd} and dual density wave~\cite{Nakano:2004cd}.
Regarding phenomenology, 
magnetic fields at considerable strength do exist in physical contexts like magnetars~\cite{Duncan:1992hi,Turolla:2015mwa,Kaspi:2017fwg} and the quark-gluon plasma~\cite{Shuryak:1978ij,Skokov:2009qp,McLerran:2013hla}. 
Also, the presence of a magnetic field may alleviate the sign problem in LQCD~\cite{Fukuda:2013ada,Brauner:2019rjg}. 

Particularly concerned with the hadronic sector are the proposed $\pi^0$ (neutral pion) domain wall~\cite{Son:2007ny} and its stacking outcome: the chiral soliton lattice (CSL)~\cite{Eto:2012qd,Brauner:2016pko}.
The conserved topological charge reduces the free energy via the WZW term and triggers a phase transition from vacuum to CSL when there is an external magnetic field exceeding the critical value $B_{\text{CSL}} = 16\pi m_\pi f_\pi^2 / (e \mu_B)$, with $m_\pi$ the effective pion mass and $f_\pi$ the pion decay constant.
Each chiral soliton carries a baryon number as the conserved charge of the GW current~\cite{Son:2007ny}.
These studies have inspired the thread that, under a strong magnetic field, the ground state of dense QCD matter may be pure pionic~\cite{Brauner:2018mpn}, contrary to the traditional baryonic picture.
Analogous CSL phases composed of $\eta'$ mesons (or $\eta$ mesons in the two-flavor case)~\cite{Huang:2017pqe,Nishimura:2020odq,Chen:2021aiq}, and their non-Abelian generalizations
involving pion degrees of freedom~\cite{Eto:2021gyy,Eto:2023rzd}, have also been proposed in rapidly rotating systems.
Furthermore, when $\eta$ mesons become physically relevant at high density in a magnetic field, two of the present authors have proposed a mixed soliton lattice comprised of both $\pi^0$ and $\eta$ mesons~\cite{Qiu:2023guy}, which features lower energy than their separated lattices, suggesting the novel possibility of a QCD quasicrystal.
In order to clarify the stability of CSLs in the non-perturbative regime, CSL ground states have been investigated in QCD-like theories such as
two-color ($N_c=2$) 
QCD and vector-like gauge theories~\cite{Brauner:2019rjg,Brauner:2019aid}, where the sign problem could be circumvented in lattice gauge theory at finite baryon density.
Also of interest are the CSL counterparts in 
supersymmetric QCD~\cite{Nitta:2024xcu}
and holographic QCD~\cite{Amano:2025iwi}.

At high density and magnetic field, there appears an intriguing phase that consists of the hybrid baryonic structure, 
domain-wall Skyrmions~\cite{Eto:2023lyo,Eto:2023wul,Amari:2024fbo,Eto:2023tuu}, which is characterized by $\pi_2(S^2) \simeq \mathbb{Z}$ and lodged on chiral solitons~\cite{Nitta:2012wi,Nitta:2012rq,Gudnason:2014nba,Gudnason:2014hsa}.
The domain-wall Skyrmion has baryon number two and hence a bosonic nature \cite{Amari:2024mip}.
At low magnetic field and high chemical potential, a Skyrmion crystal~\cite{Klebanov:1985qi} derived from the Skyrme model~\cite{Skyrme:1961vq,Zahed:1986qz} readily describe dense baryonic matter in the large $N_c$ limit. 
On top of ChPT, the Skyrme model 
incorporates an additional quartic term, namely the Skyrme term, for the stability of the Skyrmion solution, according to Derrick's theorem~\cite{Derrick:1964ww}. 
In contrast to the domain-wall Skyrmions, a Skyrmion is a single baryon with the GW current charge representing the baryon number one~\cite{Adkins:1983ya}. 
Effects of magnetic fields have been incorporated in the Skyrme model innovatively in refs.~\cite{Chen:2021vou,Chen:2023jbq}, indicating that the $\pi^0$ CSL and related configurations occupy the region of lower density and higher magnetic field in the phase diagram.
However, exactly how CSL, domain-wall Skyrmion, and Skyrmion crystal transit to each other is still an open question.
Approaching this issue, a pancake structure of the baryon in a magnetic field has been
conjectured~\cite{Son:2007ny} and under examination~\cite{Amari:2025twm}.

In realistic physical contexts, the isospin chemical potential $\mu_I$ often needs to be taken into account; for two-flavor ($N_f=2$) QCD with $u$ and $d$ quarks, the isospin chemical potential is given by the difference between chemical potentials of $u$ and $d$, i.e., $\mu_I=(\mu_u - \mu_d)/2$. Meanwhile the baryon chemical potential is defined as $\mu_B=(\mu_u + \mu_d)/2$.
Finite $\mu_{I,B}$ are pertinent in the early universe with lepton flavor asymmetry, in astrophysical objects such as neutron stars, and, of course, in heavy-ion collisions.
It has long been known that the Bose–Einstein condensation of charged pions $\pi^\pm$ occurs when $\mu_I \geq m_\pi$~\cite{Ruck:1976zt, Migdal:1978az}.
At higher $\mu_I$, a deconfined superconducting phase has been hypothesized~\cite{Son:2000xc}.
Unlike $\mu_B$, finite $\mu_I$ does not introduce a sign problem, enabling significant progress in LQCD studies~\cite{Kogut:2002tm,Kogut:2002zg,Kogut:2004zg,Brandt:2017oyy,Brandt:2022hwy,Abbott:2023coj}.
These developments have astrophysical implications, including pion stars~\cite{Brandt:2018bwq,Stashko:2023gnn,Kirichenkov:2023omy} and gravitational wave signals from pion condensates in the early universe~\cite{Vovchenko:2020crk}. 
In neutron stars where both $\mu_B$ and $\mu_I$ are finite, $\mu_I$ contributes to the nuclear symmetry energy, whose constraints remain an active research topic~\cite{Drischler:2020hwi,PREX:2021umo,Reed:2021nqk,Neill:2022psd}.
The interplay between $\mu_I$ and $\mu_B$ is rather underplayed, but it leads us to wonder novel states of matter given the $\pi^\pm$ condensate involved with dense baryonic matter. 

Motivated by the above research interests, we will study baryons in low-energy QCD with a magnetic field under the framework of ChPT at finite $\mu_B$ and $\mu_I$.
In our scenario, the relevant gauge fields are the electromagnetic $A_\mu$ and the baryonic $A_\mu^B$ involved in the WZW term.
$\mu_I$ and $\mu_B$ correspond to the zeroth components of $A_\mu$ and $A_\mu^B$, respectively.
Like in metallic type-II superconductors, Abrikosov-Nielsen-Olesen (ANO) vortices \cite{Abrikosov:1956sx,Nielsen:1973cs} are fundamental topological objects 
among the $\pi^\pm$ condensation at $\mu_I \geq m_\pi$~\cite{Ruck:1976zt, Migdal:1978az,Son:2000xc}
in the presence of a magnetic field $B$.
ANO vortices at finite $\mu_I$ and $B$ but $\mu_B = 0$ have been studied in refs.~\cite{Adhikari:2015wva,Adhikari:2018fwm,Canfora:2020uwf,Adhikari:2022cks,Gronli:2022cri}, the last of which proposed a phase diagram involving CSL and an Abrikosov vortex lattice (AVL).\footnote{Baryons were considered in ref.~\cite{Canfora:2020uwf} but without $\mu_B$ effects.}
On the other hand, refs.~\cite{Evans:2022hwr,Evans:2023hms} explored an Abrikosov vortex lattice with $\mu_B$ and $B$ but omitted $\mu_I$ effects.
The latter configurations may arise from the instability of CSL with $\pi^\pm$ condensation at large $\mu_BB$~\cite{Brauner:2016pko}.

Only very recently, ANO-like vortices with both $\mu_B$ and $\mu_I$ have been investigated in the chiral limit $m_\pi=0$ 
\cite{Qiu:2024zpg}. 
Such an ANO-like vortex is actually an ANO vortex of the charged pion 
$\pi^\pm$ confining an additional neutral pion $\pi^0$ in its core, 
yielding a $U(1)$ modulus as a superconducting cosmic string \cite{Witten:1984eb}.
$\pi^0$ varies linearly along the vortex line ($z$-axis), reducing the vortex energy via the WZW term, in the same way as the $\pi^0$ domain wall in CSL. 
The resulting vortex features lower energy than $\pi^\pm$ ANO vortex and carries a conserved baryon number, understood from the homotopy equivalent to a Skyrmion. Such a vortex is therefore dubbed a ``baryonic vortex''~\cite{Qiu:2024zpg}. 
This construction could be supported by earlier works
on vortex Skyrmions~\cite{Gudnason:2014hsa,Gudnason:2014jga,Gudnason:2016yix,Nitta:2015tua}, where twisting the $U(1)$ modulus in vortex configurations induces a Skyrmion-like baryon charge. 
The critical $\mu_B$ above which the baryonic vortex became energetically favorable over the uniform $\pi^\pm$ condensate was also identified in ref.~\cite{Qiu:2024zpg}, manifesting its significance in studies of the phase diagram.
One of the important consequences is that for sufficiently large $\mu_{\rm B}$, the tension of the baryonic vortex becomes negative, meaning that the vortex is spontaneously created in the ground state. 
However, a drawback in this analysis was that the Skyrme term is demanded to stabilize the baryonic vortex solution since there is instability at large $\mu_B$ if without the Skyrme term.

In this paper, we investigate the ANO vortex and the baryonic vortex in the {\it leading-order} ChPT with pion mass and coupling to Maxwell electrodynamics taken into account.
In contrast to the chiral limit
where, as mentioned, $\pi^0$ proves a linear function of $z$ lodged inside the $\pi^\pm$ ANO vortex core, 
when $m_\pi \neq 0$,
both $\pi^0$ and $\pi^\pm$ condense in the bulk. In other words, $\pi^0$ is no longer confined but rather has a nontrivial radial distribution and a nonvanishing condensate at infinity.
As a result, two distinct types of vortex strings can exist: an ANO-like local vortex, around which $\pi^\pm$ winds, and a global vortex, around which $\pi^0$ winds. The latter is further accompanied by a chiral soliton, a.k.a. $\pi^0$ domain wall, at the center of the vortex.
We have found that, when the baryon chemical potential $\mu_B$ exceeds a certain critical value depending on the isospin chemical potential $\mu_I$, a straight ANO vortex becomes linked with a closed $\pi^0$ string. As demonstrated in previous works~\cite{Gudnason:2020luj,Gudnason:2020qkd}, the linking number of these two types of vortices is identical to the baryon number. 
Our configuration carrying baryon number one coincides with a single Skyrmion in terms of the homotopy.
In this way, our baryonic vortex is a realization of a Skyrmion  
in a model-independent manner, i.e., from the leading order ChPT without the Skyrme term. 
Linked vortices describing a nucleon is an incarnation of Lord Kelvin's hypothesis on atoms as knotted vortices  
\cite{Thomson:1869}.
Theoretically, such a linked vortex configuration closely resembles those studied in ref.~\cite{Eto:2024hwn}, which are shown to have a non-invertible symmetry~\cite{Hidaka:2024kfx}.

This paper is organized as follows:
In Sec.~\ref{sec:chiral-Lag}, we review the chiral Lagrangian including the WZW term relevant to our study.
In Sec.~\ref{sec:ansatz}, we introduce the Ansatz for the linked baryonic vortex configuration.
In Sec.~\ref{sec:EOM}, 
we present numerical solutions that realize the linked baryonic vortex.
In Sec.~\ref{sec:vortex-phase}, 
we prescribe the phase boundary of a vortex 
between a baryonic vortex and an ANO vortex.
Finally, Sec.~\ref{sec:summary} is devoted to summary and discussion.

\section{Chiral Lagrangian with the Wess-Zumion-Witten Term}\label{sec:chiral-Lag}

The theoretical framework begins with the leading-order chiral Lagrangian
\begin{equation}
\mathcal{L}_{\text{chiral}}=\frac{f_{\pi}^{2}}{4}\left[\mathrm{Tr}\left(D^{\mu}\Sigma^{\dagger}D_{\mu}\Sigma\right)+m_{\pi}^{2}\mathrm{Tr}\left(\Sigma^{\dagger}+\Sigma-2\right)\right],
\label{eq:Lchiral}
\end{equation}
for the $SU(2)$ field parametrized as follows:
\begin{equation}
\Sigma=\sigma+i\tau_{i}\pi_{i}=\left(\begin{array}{cc}
\sigma+i\pi_{3} & -\pi_{2}+i\pi_{1}\\
\pi_{2}+i\pi_{1} & \sigma-i\pi_{3}
\end{array}\right)
\equiv\left(\begin{array}{cc}
\phi_{1} & -\phi_{2}^{\ast}\\
\phi_{2} & \phi_{1}^{\ast}
\end{array}\right),
\end{equation}
in terms of the complex scalar fields with the constraint $\det \Sigma = |\phi_1|^2 + |\phi_2|^2=1$.
One can perceive the association of $\phi_1$ with the neutral pion $\pi^0$ and $\phi_2$ with the charged pions $\pi^\pm$. 
We fix the parameter 
{$f_{\pi}=93\text{ MeV}$ and $m_{\pi}=138\text{ MeV}$. 
The covariant derivative encodes the $U(1)$ gauge field $A_{\mu}$, i.e., 
\begin{equation}
D_{\mu}\Sigma=\partial_{\mu}\Sigma-iA_{\mu}\left[Q,\Sigma\right],
\end{equation}
with the charge matrix $Q=1/6+\tau^{3}/2$. 
In our study, the isospin chemical potential $\mu_{I}$ is treated as a static and homogeneous input parameter, which can be set into the temporal component of the gauge field
$A_{0}=\mu_{I}$ effectively.  
We consider the scenario with a dynamical (though time-independent) magnetic field to be solved consistently with $\Sigma$, so the electromagnetic (EM) action is relevant:
\begin{equation}
\mathcal{L}_{\text{EM}}=\frac{1}{4 e^2}F_{\mu\nu}F^{\mu\nu},
\end{equation}
where $F_{\mu\nu}=\partial_{\mu}A_{\nu}-\partial_{\nu}A_{\mu}$.
Concerning EM interactions, 
$\mu_I$ acts as a background electric potential except in the WZW term.
The gauged WZW term originated from the triangle anomaly takes the following form in ChPT~\cite{Son:2007ny}: 
\begin{equation}
\mathcal{L}_{\text{WZW}}=\left(A_{\mu}^{B}+qA_{\mu}^{\text{EM}}\right)j_{B}^{\mu}, \label{eq:L-WZW}
\end{equation}
with $q=1/2$.
Beware $A_{\mu}^{\text{EM}}=A_{\mu}-\delta_{\mu0}\mu_{I}$ is the
pure EM field with isospin chemical potential subtracted. 
We adopt a distinct notation for the effective baryon gauge field $A_{\mu}^{B}=\delta_{\mu0}\mu_{B}$ to capture the effect of the baryon chemical potential.

Hereby, we comment on the symmetry of the theory. If the isospin chemical potential $\mu_I$, the pion mass $m_\pi$, and the EM interaction were absent, the Lagrangian would have an $SU(2)_L\times SU(2)_R$ chiral symmetry 
\begin{align}
    \Sigma\to U_L^\dagger \, \Sigma \,U_R, \quad U_{L,R} \in SU(2)_{L,R},
\end{align}
spontaneously broken down to the diagonal $SU(2)_V$ with $U_L=U_R$ due to the vacuum expectation value (VEV) 
chosen as
$\Sigma=1_{2\times2}$,
leading to the target space $SU(2)\simeq S^3$.
Turning on the $m_\pi$-term in the Lagrangian~\eqref{eq:Lchiral} term with $m_\pi$} explicitly breaks the chiral symmetry $SU(2)_L\times SU(2)_R$ into $SU(2)_V$, giving mass to the pions. 

On the other hand, introducing the electromagnetic $U(1)_\mathrm{EM}$ gauge field explicitly breaks the chiral symmetry $SU(2)_L\times SU(2)_R$ into $U(1)_L\times U(1)_R$,
generated by $\tau^3$:
\begin{equation}
    \Sigma \to e^{-i\alpha_L \tau_3} \Sigma e^{i\alpha_R \tau_3},
\end{equation}
which could be transcribed as 
\begin{equation}
\phi_1 \to e^{-i (\alpha_L-\alpha_R)}\phi_1 , 
\quad \phi_2 \to e^{  i (\alpha_L+\alpha_R)}\phi_2 ,    
\end{equation}
with $ e^{-i (\alpha_L-\alpha_R)} \in U(1)_{3,A}$ and 
$e^{  i (\alpha_L+\alpha_R)} \in U(1)_{\rm EM}$. 
The four $U(1)$ symmetries are related as
\begin{align}
    \frac{U(1)_L \times U(1)_R}{{\mathbb Z}_2}
    \simeq 
    U(1)_{3, A} 
    \times U(1)_{\rm EM} ,
    \label{eq:U(1)xU(1)}
\end{align}
where ${\mathbb Z}_2$ in the denominator of the left hand side is needed because the action of $U(1)_L \times U(1)_R$ at $\alpha_L=\alpha_R=\pi$ does not act on $\phi_1$ and $\phi_2$ so it should be excluded. 
With finite pion mass $m_\pi$, 
the $U(1)_{3, A}$ symmetry is explicitly broken, 
while in the chiral limit it is intact.

Let us consider the spontaneous breakings of the symmetries in eq.~(\ref{eq:U(1)xU(1)}) with the EM interaction. 
In general, when $\phi_1$ ($\phi_2$) develops a VEV, 
the $U(1)_{3, A}$ ($U(1)_{\rm EM}$) symmetry is spontaneously broken. 
Which field develops a VEV depends on the parameters.
In the absence of $\mu_I$,
the symmetries in eq.~\eqref{eq:U(1)xU(1)} are spontaneously broken as $U(1)_\mathrm{EM}\times U(1)_{3,A} \to U(1)_\mathrm{EM}$ since 
$\phi_1$ features the VEV while $\phi_2$ does not.
Now we consider the effect of $\mu_I$ on eq.~\eqref{eq:U(1)xU(1)}.
In the chiral limit at finite $\mu_I$, $\phi_2$ develops a VEV but $\phi_1$ does not,  so the symmetry breaking is $U(1)_\mathrm{EM}\times U(1)_{3,A} \to U(1)_{3,A}$, which is the case studied in Ref.~\cite{Qiu:2024zpg}.
By contrast, when finite $m_\pi$ is taken into account, the $U(1)_{3,A}$ is broken both spontaneously and explicitly. 
Meanwhile, the $U(1)_\mathrm{EM}$ is broken only when $\mu_I>m_\pi$, in which case $\phi_1$ has VEV $m_{\pi}^{2}/\mu_{I}^{2}$
while $\phi_2$ has VEV $\sqrt{1-m_{\pi}^{4}/\mu_{I}^{4}}$.

To comprehend the topological aspect of the theory, we delve into the detailed structure
of $j_{B}^{\mu}$. To proceed, we first summon the right hand
and left hand $SU(2)$ current forms:
\begin{equation}
l=\Sigma^{\dagger}d\Sigma=\left(\begin{array}{cc}
l_{11} & l_{12}\\
-l_{12}^{\ast} & l_{11}^{\ast}
\end{array}\right),\quad r=\Sigma d\Sigma^{\dagger}=\left(\begin{array}{cc}
r_{11} & r_{12}\\
-r_{12}^{\ast} & r_{11}^{\ast}
\end{array}\right).
\end{equation}
with the notations:
\begin{align}
l_{11} & =\phi_{1}^{\ast}d\phi_{1}+\phi_{2}^{\ast}d\phi_{2},\quad l_{12}=-\phi_{1}^{\ast}d\phi_{2}^{\ast}+\phi_{2}^{\ast}d\phi_{1}^{\ast},\nonumber \\
r_{11} & =\phi_{1}d\phi_{1}^{\ast}+\phi_{2}^{\ast}d\phi_{2},\quad r_{12}=\phi_{1}d\phi_{2}^{\ast}-\phi_{2}^{\ast}d\phi_{1}.
\end{align}
Then we readily consider the baryon current~\cite{Son:2007ny} 
\begin{equation}
j_{B}=\star\frac{1}{24\pi^{2}}\mathrm{Tr}\left\{ l\wedge l\wedge l+3iQd\left[A\wedge\left(l-r\right)\right]\right\} .
\end{equation}
Among it, the first term in the bracket,
\begin{equation}
\mathrm{Tr}\left(l\wedge l\wedge l\right)=3\left[d\phi_{1}\wedge d\phi_{2}\wedge\left(\phi_{1}^{\ast}d\phi_{2}^{\ast}-\phi_{2}^{\ast}d\phi_{1}^{\ast}\right)+\mathrm{h.c.}\right],
\end{equation}
is topologically preserved, with its contribution to $j_{B}^{0}$
featuring a spatial integration that amounts to an integer governed
by $\pi_{3}(S^{3})\simeq \mathbb{Z}$. On top of that, the contribution that explicitly depends on the gauge field takes the following form
\begin{align}
\mathrm{Tr}\left\{ 3iQd\left[A\wedge\left(l-r\right)\right]\right\}  & =3id\left[A\wedge\left(\phi_{1}^{\ast}d\phi_{1}-\mathrm{h.c.}\right)\right]\nonumber \\
 & =3i\left[dA\wedge\left(\phi_{1}^{\ast}d\phi_{1}-\mathrm{h.c.}\right)+2Ad\phi_{1}\wedge d\phi_{1}^{\ast}\right].
\end{align}
It functions as a covariant generalization of the $j_{B}$. With or
without the $A_{\mu}$ related portion, the baryon number is always
conserved as it should be,
whereas we would see in what follows that the $j_{B}$ does influence the EOM of the
magnetic field and the WZW term does play a role in determining the
ground state configuration.

\section{Vortex Ansatz and Boundary Conditions}\label{sec:ansatz}
To begin with, we reiterate the gist of crafting our novel Ansatz: 1. $\pi^\pm$ local (gauged) vortex to accommodate a magnetic field, 2. $\pi^0$ winding to form a nontrivial baryon charge. 
Such an idea is investigated in ref.~\cite{Qiu:2024zpg} for the case of the chiral limit $m_{\pi}=0$ by conceiving a vortex-Skyrmion 
~\cite{Gudnason:2014hsa,Gudnason:2014jga,Gudnason:2016yix,Nitta:2015tua} with  $\phi_2=|\phi_2|\exp(in\varphi)$ and $\phi_{1}=|\phi_1|\exp(ikz)$ where the magnitudes $|\phi_{1,2}|$ are functions of the polar radius $\rho$ while the phases depend on the azimuthal angle $\varphi$ and the longitudinal coordinate $z$, in terms of the cylindrical coordinates $\left(\rho,\varphi,z\right)$. 
Obviously, $n$ means the winding number of the $\pi^\pm$ local (gauged) vortex, which amounts to one if we consider a single vortex. This vortex stretches along the $z$-axis with $\pi^0$ winding inside its core in the same manner as CSL in the chiral limit, i.e., a Bloch-like wave with wavevector $k$ propogating in the periodic longitudinal direction. In each period, we demonstrated that the baryon number, calculated by integrating $j_B^0$, is one. Thus, the number of periods is nothing but the baryon number, and we dubbed the vortex-Skymrion a baryonic vortex. In the present study, we extend this scenario to a massive version.
However, it will turn out that the topological structure of the configuration in the massive case is very different 
from that in the chiral limit.
To this end, we shall, in the first place, pay attention to the symmetry breaking pattern that is also divided.

As mentioned in Sec.~\ref{sec:chiral-Lag}, in the presence of the EM gauge interaction, the symmetry is $U(1)_{\rm EM}\times U(1)_{3,\rm A}$. 
Among it, the $U(1)_{3,\rm A}$ symmetry is spontaneously broken 
when $\mu_I=0$, 
making a global vortex of $\pi^0$ possible. 
The vortex line is attached by a $\pi^0$ domain wall,  namely a chiral soliton, for $m_\pi \neq 0$.
This phenomenon will not 
change even if a finite $\mu_I$ is introduced yet with $\mu_I<m_\pi$, 
in which case $\phi_2$ does not obtain the VEV, so that the 
symmetry breaking remains the same, 
allowing only the global vortex.
Meanwhile, if $\mu_I>m_\pi\neq0$, the $U(1)_{\rm EM}$ symmetry is spontaneously broken due to the VEV of $\phi_2$, i.e., the $\pi^\pm$ condensation, 
which allows a local $\pi^\pm$ vortex to coexist with the global $\pi^0$ vortex.
Based on such understanding, we target a single local vortex-string along the $z$-axis that is eventually linked with a closed $\pi^0$ vortex-string and intersected by the domain wall bounded by the $\pi^0$ string.
The scenario is distinguished from that in ref.~\cite{Qiu:2024zpg}, where a finite $\mu_I$ with $m_\pi=0$ prohibits the $\pi^0$ condensate and the possibility of global vortices.
As a result, our target in the present paper is valid only for $\mu_I>m_\pi$. For such a scenario, the $\pi^0$ is not confined in the vortex core but rather forms a global vortex which features nonzero VEV at infinity.

Still, we consider the solution with cylindrical symmetry apt for an axial magnetic field.  Accordingly the gauge field $\boldsymbol{A}=A_{\varphi}\left(\rho,z\right)\hat{\varphi}$ is set up along the azimuthal unit vector $\hat{\varphi}$, which yields
\begin{equation}
dA=\partial_{\rho}ad\rho\wedge d\varphi-\partial_{z}ad\varphi\wedge dz,
\end{equation}
with $a\equiv\rho A_{\varphi}$ a notation for convenience.
Note that $\mu_{I}$ has no spacetime dependence so there is no effective
electric field involved, consistent with the nature of a static configuration.
The $\varphi$ dependence of fields is
entirely attributed to
\begin{equation}
\phi_{2}=f\left(\rho,z\right)e^{i\varphi},
\label{eq:ANO}
\end{equation}
while $\phi_{1}\left(\rho,z\right)$ and $\phi_{1}^{\ast}\left(\rho,z\right)$
are functions of only $\rho$ and $z$, consistent with the cylindrical symmetry.
Due to the regularity at $\rho=0$,
we have
\begin{align}
    f(0,z)=a(0,z)=\left.\partial_\rho\phi_1\right|_{\rho=0}=0 \, . \label{eq:regular}
\end{align}
$\phi_2$ in eq.~\eqref{eq:ANO} alone is reminiscent of a $\pi^\pm$ ANO vortex. 
Now it is further associated with the $\pi^0$ to form a vortex with a nontrivial baryon number.

We anatomize the structure of $j_{B}^{0}$ to tailor the boundary
conditions that guarantee a topologically conserved baryon number.
For a static vortex, time dependence is shut down. In this case, the $j_{B}$ in terms of our Ansatz is spell out as: 
\begin{align}
\mathrm{Tr}\left(l\wedge l\wedge l\right)=-6\mathrm{Im} & \bigg[\partial_{z}\left(f^{2}\right)\phi_{1}^{\ast}\partial_{\rho}\phi_{1}-\partial_{\rho}\left( f^{2}\right)\phi_{1}^{\ast}\partial_{z}\phi_{1}\nonumber \\
 & +2 f^{2}\partial_{\rho}\phi_{1}^{\ast}\partial_{z}\phi_{1}\bigg]d\rho\wedge d\varphi\wedge dz,\\
\mathrm{Tr}\left\{ 3iQd\left[A\wedge\left(l-r\right)\right]\right\} =-6\mathrm{Im} & \bigg\{\bigg[\left(\partial_{\rho}a\right)\phi_{1}^{\ast}\partial_{z}\phi_{1}-\left(\partial_{z}a\right)\phi_{1}^{\ast}\partial_{\rho}\phi_{1}\nonumber \\
 & +2a\partial_{\rho}\phi_{1}^{\ast}\partial_{z}\phi_{1}\bigg]d\rho\wedge d\varphi\wedge dz\nonumber \\
 & +2\mu_{I}\left(\partial_{\rho}\phi_{1}^{\ast}\partial_{z}\phi_{1}\right)dt\wedge d\rho\wedge dz\bigg\}.
\end{align}
We further introduce the parametrization 
\begin{equation}
\phi_{1}=\cos\alpha(\rho,z)\exp\left(i\beta(\rho,z)\right),\quad |\phi_2| = f(\rho,z) = \sin\alpha(\rho,z),
\label{eq:defphi}
\end{equation}
with which one can reach a concise expression 
\begin{equation}
j_{B}=\star \frac{-1}{4\pi^{2}}
\left\{
d\left[\cos^{2}\alpha\left(1+a\right)\right]\wedge d\varphi\wedge d\beta
+d\left(\mu_I \cos^{2}\alpha \right)\wedge d\beta \wedge dt
\right\}.
\end{equation}
After performing the Hodge star, 
we arrive at the expression of the baryon charge density:
\begin{equation}
j_{B}^{0}=\frac{1}{4\pi^{2}\rho}\left\{ \partial_{\rho}\left[\cos^{2}\alpha\left(1+a\right)\right]\partial_{z}\beta-\left(\rho\leftrightarrow z\right)\right\} ,
\label{eq:jB0}
\end{equation}
The spatial integration yields
\begin{align}
\int d^{3}xj_{B}^{0}=\frac{1}{2\pi} & \bigg\{\int dz\cdot\left[\cos^{2}\alpha\left(1+a\right)\partial_{z}\beta\right]\bigg|_{\rho=0}^{\rho=\infty}\nonumber \\
 & -\int d\rho\cdot\left[\cos^{2}\alpha\left(1+a\right)\partial_{\rho}\beta\right]\bigg|_{z=-L/2}^{z=L/2}\bigg\}.\label{eq:jdetail}
\end{align}

We conceive a single vortex string
occupying a fixed total longitudinal length $L$.
Then the integration reduces to 
\begin{equation}
\int d^{3}xj_{B}^{0}=\frac{1}{2\pi}\int_{-L/2}^{L/2}dz\cdot\left[\cos^{2}\alpha\left(1+a\right)\partial_{z}\beta\right]\bigg|_{\rho=0}^{\rho=\infty},\label{eq:jdetail2}
\end{equation}
where the second line of eq.~\eqref{eq:jdetail} vanishes due to a periodic boundary condition at $z=\pm L/2$:
\begin{equation}
X\left(z+L\right)=X\left(z\right);\quad X=\phi_{1},\,\phi_{1}^{\ast},\,f,\,a.
\label{eq:period}
\end{equation}
since the two ends of $z$-axis are identified.
We proceed to examine
boundary conditions at $\rho=0$ and $\infty$. At infinity, the finiteness of the total energy requires a pure gauge field  
\begin{equation}
a\left(\infty,z\right)=-1,\label{eq:bc2}
\end{equation}
as well as an unchanging phase $\partial_{z}\beta\big|_{\rho=\infty}=0$. Hence the nonzero
contribution to the baryon number comes solely from the integration along
the $z$-axis at $\rho=0$. 
From the regularity condition \eqref{eq:regular}, 
we have
\begin{equation}
\int d^{3}xj_{B}^{0}=-\frac{1}{2\pi}\int_{-L/2}^{L/2}dz\,\partial_{z}\beta(0,z)\, .
\end{equation}
Therefore, the baryon number is equal to the winding number of $\beta$ along the $z$ axis.

As the simplest case, we here demand that our configuration has baryon number unity, $\int d^3x \, j_B^0=1$.
Such a configuration of $\beta(0,z)$ is given by, e.g.,
a single sine-Gordon kink
\begin{equation}
\phi_{1}\left(0,z\right)=\exp\left\{ i\cdot4\arctan\left[\exp(-m_{\pi}z)\right]\right\} \, .
\label{eq:bc5}
\end{equation}
In fact, this configuration \eqref{eq:bc5} is a solution of the EOM of $\phi_1$ at $\rho=0$ since the Lagrangian reduces to that in the sine-Gordon model with the condition $a(0,z)=f(0,z)=0$.\footnote{
It is straightforward to generalize $\phi_1$ to have an integer baryon number $N$,
which means that $\beta$ has a winding number $N$.
Such a configuration is given by arranging $N$ sine-Gordon kinks along $z$ axis,
$
\phi_{1}\left(0,z\right)=\exp\left\{ i\cdot2\arccos\left[\mathrm{sn}\left(m_{\pi}z,\kappa \right)\right]\right\} \, ,
$
where $\mathrm{sn}$ stands for the Jacobi elliptic sine and the elliptic
modulus $\kappa$ is related to a longitudinal unit period $l\equiv L/N$ by $l=2\kappa K\left(\kappa \right)/m_{\pi}$
with $K\left(\kappa \right)$ the complete elliptic integral of the first kind. 
Familiar readers may realize such an expression is nothing but the CSL solution established in ref.~\cite{Brauner:2016pko}.
Now we just implement it inside the vortex center. 
In this paper, we concentrate on the simple yet important case 
of $N=1$ soliton vortex in an infinitely large system, i.e.,
$l=L\rightarrow\infty$.}
The physical interpretation of this configuration is a single vortex carrying a unit baryon number.
Delving into such a case is relevant in the sense that if a phase transition from ANO vortices (of pure $\pi^\pm$) into baryonic vortices occurs, it shall start from such a single-baryon configuration.
In other words, the single-baryon vortex sets the boundary on the phase diagram,
between an AVL and a baryonic vortex lattice.

So far, the baryon number is guaranteed, prescribing
the behavior of $\Sigma$ at $\rho=0$. The last step in fixing boundary conditions is to incorporate the pion condensate induced by $\mu_I$ at $\rho\rightarrow\infty$. As is well known, for $\mu_{I}>m_{\pi}$,
the magnitude of $\pi^\pm$ condensate is
\begin{equation}
f\left(\infty,z\right)=\sqrt{1-m_{\pi}^{4}/\mu_{I}^{4}}.\label{eq:bc6}
\end{equation}
Meanwhile, at $\rho\rightarrow\infty$ the neutral pion phase $\beta$ (seen in eq.~\eqref{eq:defphi}) remains constant.  
The specific value of $\beta\left(\rho=\infty\right)$
is irrelevant in that field configurations with different values of
$\beta$ relate to each other by a trivial redefinition. We therefore
stipulate without loss of generality
\begin{equation}
\phi_{1}\left(\infty,z\right)=\phi_{1}^{\ast}\left(\infty,z\right)=m_{\pi}^{2}/\mu_{I}^{2}.\label{eq:bc7}
\end{equation}
In summary, 
we will look for a configuration satisfying the boundary conditions eqs.~\eqref{eq:regular}, \eqref{eq:bc2}, \eqref{eq:bc6}, and \eqref{eq:bc7} 
with $\beta$ carrying a unit winding number along $z$ axis like eq.~\eqref{eq:bc5}.
We highlight that the nonzero vacuum condensate of $\pi^0$ seen from eq.~\eqref{eq:bc7}, 
when considered together with the winding at $\rho=0$, i.e., 
eq.~\eqref{eq:bc5}, implies a linking between $\pi^0$ and $\pi^\pm$. Because the phase of $\phi_1$ winds $2\pi$ following the contour of the half $\rho z$-plane, there shall exist a zero of $|\phi_1|$ as 
the intersection of the $\rho z$-plane and
the center of a closed $\pi^0$ vortex string. Meanwhile, the $\pi^\pm$ vortex center lies along the $z$-axis. The two vortices link to each other, forming the conserved baryon number $\int{j_B^0\,d^3x}$ as the linking number \cite{Gudnason:2020luj,Gudnason:2020qkd}. Indeed, in the subsequent section, we will show that the solution profile meets this expectation. Viewed as a whole, the configuration is interpreted as a ``baryonic vortex with linking number''. 

\section{Baryonic Vortex Solutions with Linking Number}\label{sec:EOM}

The EOMs are derived from the variational principle of the Hamiltonian.
The $\mathcal{O}\left(p^{2}\right)$ chiral Lagrangian yields the following Hamiltonian density 
\begin{align}
\mathcal{H}_{\text{chiral}}=\frac{f_{\pi}^{2}}{2} & \bigg\{\partial_{\rho}\phi_{1}\partial_{\rho}\phi_{1}^{\ast}+\partial_{z}\phi_{1}\partial_{z}\phi_{1}^{\ast}+\left(\partial_{\rho}f\right)^{2}+\left(\partial_{z}f\right)^{2}\nonumber \\
 & +f^{2}\left[\left(\frac{1+a}{\rho}\right)^{2}-\mu_{I}^{2}\right]+m_{\pi}^{2}\left(2-\phi_{1}-\phi_{1}^{\ast}\right)\bigg\}.
\end{align}
The $\mu_B$ does not enter the EOMs since the $j_{B}^0$ yields a conserved charge after the integral, which can not be varied. 
One can also perceive this point from the concrete form of the energy density contributed by WZW term [see eqs.~\eqref{eq:L-WZW} and  \eqref{eq:jB0}]
\begin{equation}
\mathcal{H}_{\text{WZW}}=-\mu_{B}j_{B}^{0}-\frac{q\mu_{I}}{2\pi^{2}}\mathrm{Im}\left(\partial_{\rho}\phi_{1}^{\ast}\partial_{z}\phi_{1}\right)A_{\varphi}. \label{eq:WZW}
\end{equation}
It is noteworthy that the WZW term does affect the EOM via $\mu_I$, 
which can be plainly seen if the EOM of $A_\varphi$ is written in the form of the
Maxwell equation
\begin{equation}
\partial_{\mu}F^{\mu\varphi}=j_{Q}^{\varphi}\equiv qj_{B}^{\varphi}+j_{I}^{\varphi}.\label{eq:maxwell}
\end{equation}
The right hand side of eq.~\eqref{eq:maxwell} is the survival component of the electric current, composed
of the baryon current 
\begin{equation}
j_{B}^{\varphi}=\frac{\mu_{I}}{2\pi^{2}}\mathrm{Im}\left(\partial_{\rho}\phi_{1}^{\ast}\partial_{z}\phi_{1}\right),
\end{equation}
and the isospin current
\begin{equation}
j_{I}^{\varphi}=-\frac{\delta\mathcal{L}_{\text{chiral}}}{\delta A_{\varphi}}=f_{\pi}^{2}\left(\frac{1+a}{\rho}\right)f^2.
\end{equation}
The left hand side of eq.~\eqref{eq:maxwell} can be derived from the EM sector 
\begin{equation}
\mathcal{H}_{\text{EM}}=\frac{1}{2e^2}\left[\frac{1}{\rho}\partial_{\rho}\left(\rho A_{\varphi}\right)\right]^{2}.
\end{equation}
Now we have explicated all parts of the total energy 
\begin{equation}
H=2\pi\int\rho \left(\mathcal{H}_{\text{chiral}}+\mathcal{H}_{\text{EM}}+\mathcal{H}_{\text{WZW}}\right) d\rho dz,
\end{equation}
which is to be operated with the action principle.

For numerical purposes, we need to rescale the space coordinate and the gauge field as 
\begin{equation}
\tilde{x}^\mu=f_{\pi}x^\mu,\quad\tilde{A}_{\mu}=A_{\mu}/f_{\pi},
\end{equation}
so that we can deal with all dimensionless quantities after further defining 
\begin{equation}
\tilde{H}=H/f_{\pi},\quad\tilde{m}_{\pi}=m_{\pi}/f_{\pi},\quad \left(\tilde{\mu}_I=\mu_I/f_\pi\right).
\end{equation}
Hereafter, we abbreviate the tilde and use rescaled quantities by default. 
We lay out the EOMs derived from $\delta H = 0$ as follows: 
\begin{align}
\phi_{1}:\quad & \partial_{\rho}\left(\rho\partial_{\rho}\phi_{1}^{\ast}\right)+\partial_{z}\left(\rho\partial_{z}\phi_{1}^{\ast}\right)=-m_{\pi}^{2}\\
\phi_{1}^{\ast}:\quad & \partial_{\rho}\left(\rho\partial_{\rho}\phi_{1}\right)+\partial_{z}\left(\rho\partial_{z}\phi_{1}\right)=-m_{\pi}^{2}\\
\phi:\quad & \partial_{\rho}\left(\rho\partial_{\rho}\phi\right)+\partial_{z}\left(\rho\partial_{z}\phi\right)=\rho f \left[\left(\frac{1+a}{\rho}\right)^{2}-\mu_{I}^{2}\right]\\
A_{\varphi}:\quad & \partial_{\rho}\left(\frac{1}{\rho}\partial_{\rho}a\right)+\partial_{z}\left(\frac{1}{\rho}\partial_{z}a\right)=
e^2\left[\phi^{2}\frac{1+a}{\rho}-\frac{q\mu_{I}}{2\pi^{2}}\text{Im}\left(\partial_{\rho}\phi_{1}^{\ast}\partial_{z}\phi_{1}\right)\right] \, .
\end{align}

When solving the EOMs numerically,
first we need a constraint to ensure $|\phi_1|^2+|\phi_2|^2=1$,
which is approximately realized by adding a term $\lambda \, m_\pi^2 f_\pi^2 (|\phi_1|^2+|\phi_2|^2-1)^2$ to $\mathcal{H}_\mathrm{chiral}$ (and corresponding terms to the EOMs above) with a sufficiently large $\lambda=2000$.\footnote{
Due to this term, the VEV of $f$ is slightly modified into $f(\infty,z)^2=1-m_\pi^4/\mu_I^4 + \mu_I^2/(4\lambda m_\pi^2)$. 
However, it is a minor technical modification that would not affect our results significantly.
}
Then we discretize the $(\rho,z)$-space into a two-dimensional lattice with different lattice spacings in each direction: $\Delta z = 0.1 f_\pi^{-1}$ and $\Delta \rho = 0.01 f_\pi^{-1}$.
The simulation box size is taken as $0\leq \rho \leq 15f_{\pi}^{-1}$ and $-10f_{\pi}^{-1}\leq z \leq 10f_{\pi}^{-1}$, which are sufficiently large compared to the typical length scale of the configuration.
We then adopted the relaxation method, 
i.e., start from an appropriate initial configuration and update the field variables iteratively with a step change in time $\Delta t = 2\times 10^{-6} \, f_\pi^{-1}$ until they converge to the solution of the EOMs.

We solved the EOMs at different input values of $\mu_I$. For $\mu_I \lesssim 3.5m_\pi$ we found no stable solution within our numerical approach. However, that does not rule out the possibility of the existence of solutions. 
Rigorously speaking, we do not know what happens to the baryonic vortex for $\mu_I \lesssim 3.5m_\pi$.  

It has been known that for $\mu_I \geq m_\pi$ the ground state is a homogeneous charged pion condensate. 
ANO vortex would arise from such a condensate if an external magnetic field is applied.
Along this line, in the current work, we would not repeat what had already been explored in ref.~\cite{Gronli:2022cri}. Instead, we would assume an already existing ANO vortex without considering the background magnetic field, which excites the vortex precedingly.
What our results show originally is: when $\mu_\text{B}$ is turned on, the ANO vortex could transit into a vortex with the linking number that has the physical meaning as the baryon number, thus a ``baryonic vortex''.

\begin{figure}[tbp]
 \centering \includegraphics[width=0.7\textwidth]{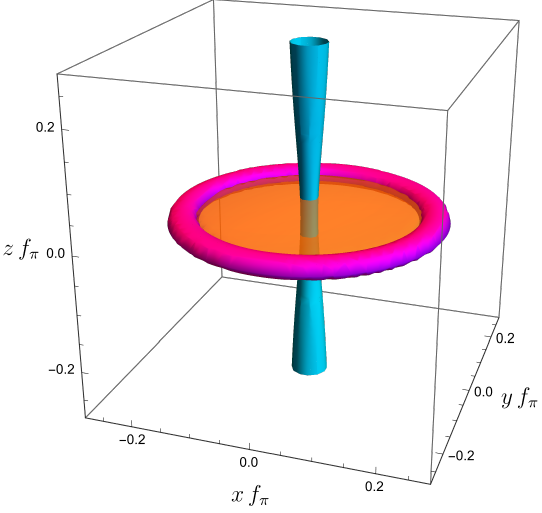}
 \caption{A 3D configuration of a linked baryonic vortex. The parameter choice is $\mu_I/m_\pi=4.0$. The light blue, purple and orange surfaces are defined by the conditions 
$|\phi_2|/|\phi_2(\infty,z)|<0.25$, 
$|\phi_1|/|\phi_1(\infty,z)|<0.25$,  
and 
$\pi - 0.5 <\beta< \pi + 0.5 $, 
respectively. 
They denote the ANO-like vortex string of the $\pi^\pm$ string, 
closed $\pi^0$ vortex string, 
and $\pi^0$ domain wall (chiral soliton), respectively. 
}
 \label{fig:3D-link}
\end{figure}
\begin{figure}[tbp]
 \centering
 \begin{tabular}{cc}
  \includegraphics[width=0.47\textwidth]{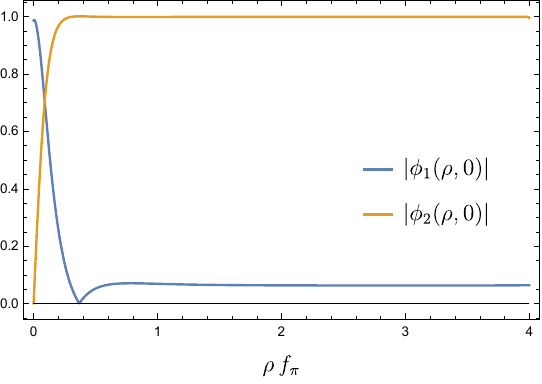}    &  
  \includegraphics[width=0.47\textwidth]{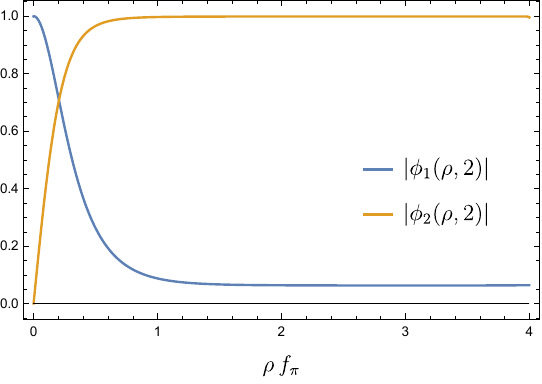}\\
  (a) & (b)\\
  \includegraphics[width=0.47\textwidth]{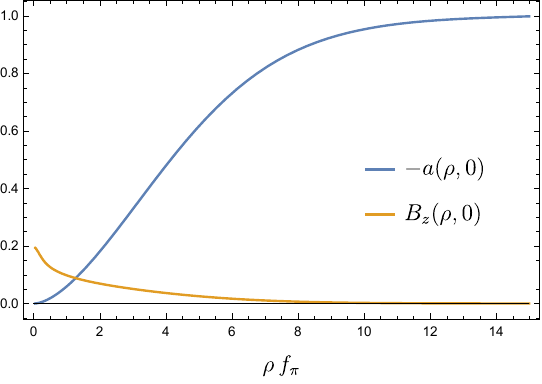}   
   &
  \includegraphics[width=0.47\textwidth]{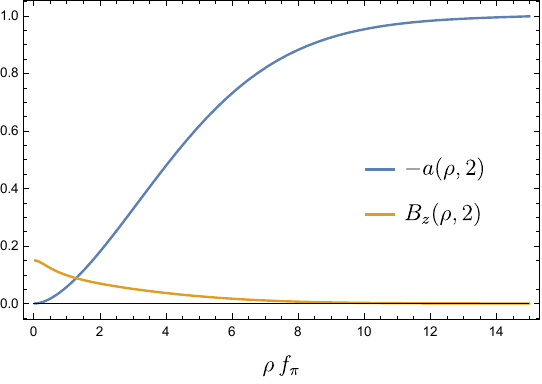}\\
  (c) & (d)
 \end{tabular}
 \caption{
Plots of the profiles of the linking solution with a benchmark case $\mu_I/m_\pi=4.0$.
(a) and (b): 
The profiles of $|\phi_1|$ and $|\phi_2|$
depending on the radial coordinate $\rho$ 
on the planes of $z=0$ and $z=2$, respectively.
(c) and (d):
The profiles of the gauge field $a$ and the magnetic 
fields $B_z$ depending on $\rho$, 
corresponding to (a) and (b). 
$\phi_2$ vanishes along the $z$ axis $\rho=0$,
which is the core of the local vortex string of $\pi^\pm$, accommodating the localized magnetic field $B_z$.
On the other hand, there are zeros of $\phi_1$ 
at $\rho\simeq 0.4/f_\pi$ and $z=0$ [(a)], corresponding to the core of the closed global vortex string of $\pi_0$, 
while there is no zero at $z=2$ [(b)], as at any $z\neq0$. 
}
 \label{fig:profile}
\end{figure}

\begin{figure}[tbp]
 \centering
 \includegraphics[width=0.65\textwidth]{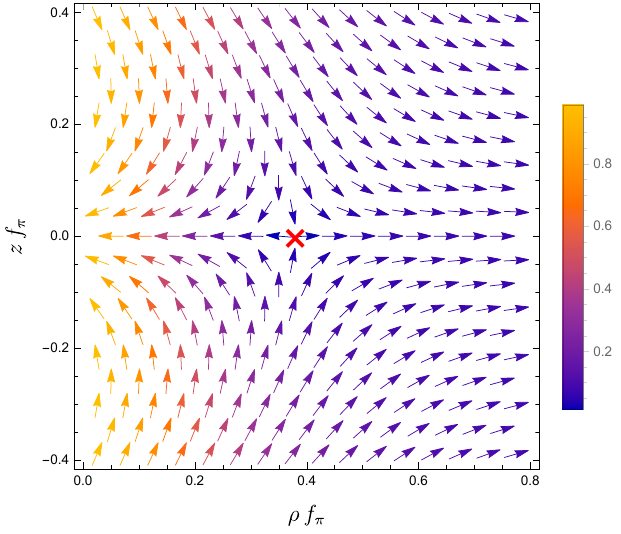}
 \caption{
The phase and amplitude of the neutral pion $\phi_1$ 
in the $(\rho,z)$ plane at $\mu_I=4.0m_\pi$.
Vectors correspond to $(\mathrm{Re}\, \phi_1,\mathrm{Im}\, \phi_1)$. The color corresponds to the amplitude $|\phi_1|$. 
The cross denotes the position of the $\pi^0$ vortex core.
}
 \label{fig:phases}
\end{figure}
We exemplify the solution of our baryonic vortex with its profile at $\mu_I=4m_\pi$. 
Fig.~\ref{fig:3D-link} shows a 3D plot of the whole configuration. 
The ANO vortex string of the $\pi^\pm$ winding is illustrated by
the light blue region: 
$|\phi_2|/|\phi_2(\infty,z)|<0.25$.
The closed $\pi^0$ vortex is illustrated by the purple region: 
$|\phi_1|/|\phi_1(\infty,z)|<0.25$. 
At last, the $\pi^0$ domain wall (chiral soliton) is illustrated by the orange region.

Fig.~\ref{fig:profile} shows the magnitudes of the $\phi_1$, $\phi_2$, $a(\rho,z)$, and the magnetic field $B_z\equiv -(\partial_\rho a)/\rho$. 
The left panel shows those at the $z=0$ slice, in which one can observe that the amplitude $|\phi_1|$ of $\pi^0$ reaches zero at $\rho \simeq 0.38 f_{\pi}^{-1}$ 
implying the core of the neutral pion vortex ring, the purple ring in fig.~\ref{fig:3D-link}. 
The right panel shows the profile functions at the $z=2 f_\pi^{-1}$ slice, 
in which there are no zeros of $|\phi_1|$.

A plot of the phase of neutral pion 
as a vector $(\mathrm{Re}\, \phi_1,\mathrm{Im}\, \phi_1)$ 
in the $(\rho,z)$ plane is shown 
in fig.~\ref{fig:phases}. 
One can clearly see the center of the neutral pion vortex 
denoted by the cross around at $(\rho,z)\simeq (0.38f_\pi^{-1},0)$. 

\section{Vortex Phase Transition}\label{sec:vortex-phase}

\begin{figure}[tbp]
 \centering
 \includegraphics[width=0.7\textwidth]{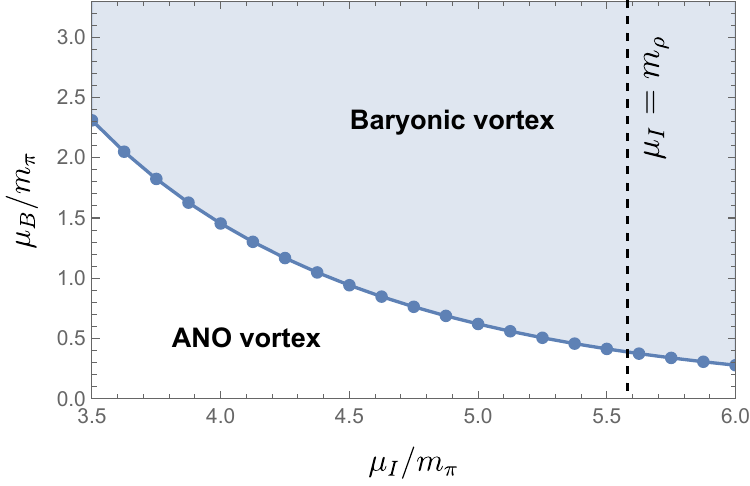}
 \caption{
 A plot of the vortex phase diagram.
 The blue region corresponds to the baryonic vortex phase, 
 in which baryonic vortices have lower energy than ANO vortices, while the white region corresponds to a conventional phase. 
 The vertical dashed line corresponds to the $\mu_I$ equal to the mass of $\rho$ meson $m_\rho$. 
 For $\mu_I>m_\rho$, the description based on ChPT breaks down.}
 \label{fig:phase-diagram}
\end{figure}

Targeting the role played by the baryonic vortex on the phase diagram,
we evaluate the free energy by applying the numerical solutions of $\Sigma$ and $A_\varphi$, which have been explained above.
$\mu_B$ contributes to the free energy via the WZW term, amounting to $-\mu_B N$ as expected for a chemical potential. 
Then the total energy
would become lower than that of a single ANO vortex at sufficiently large $\mu_B$, i.e.,
\begin{equation}
    \mu_B>\mu_c=\int d^3x \left(\mathcal{H}_\text{chiral}+\mathcal{H}_\text{EM}-j_B^\varphi A_\varphi\right);\quad H<0.
\end{equation}
Such a critical $\mu_c$ can be evaluated after applying the numerical solutions of $\Sigma$ and $A_\mu$.
For each value of $\mu_I$, upon the existence of the baryonic vortex solution, there is a corresponding $\mu_c$ above which an ANO vortex turns to a baryonic vortex, marking the onset of the nucleation when $\mu_B$ is increased from below.
The $\mu_c$ as a function of $\mu_I$ (or vice versa) is plotted in the phase diagram in fig.~\ref{fig:phase-diagram}, standing as the boundary between phases of the ANO vortex and the baryonic vortex.

The scenario bears similarity to the transition between CSL and domain wall Skyrmion phases in a magnetic field (rather than $\mu_I$) \cite{Eto:2023lyo,Eto:2023wul,Amari:2024fbo,Eto:2023tuu}, which in view of a single soliton is the growth of a baby Skyrmion on top of a uniform $\pi^0$ domain wall. The difference consists in the fact that the ANO vortex does not carry the baryon number that our baryonic vortex carries. Thus, the baryon number can be regarded as an order parameter for the phase transition we established.
\footnote{
A similar topological phase transition was discussed in a chiral magnet as the transition between a uniform ferromagnet state and the CSL in condensed matter physics~\cite{Amari:2024jxx}.
}

Remarkably, as shown in fig.~\ref{fig:phase-diagram}, the $\mu_c$ we found for the baryonic vortex proves smaller than the critical $\mu_B$ to excite a nucleon in the bulk \cite{Amari:2025twm}, which is nothing but the nucleon mass $m_N\simeq 940 \text{MeV}$ if there is no external field. $\mu_c < m_N$ demonstrates the influence of $\mu_I$ on the ground state, i.e., the $\pi^\pm$ condensate is energetically preferred in the first place, and when $\mu_B$ is increased from zero in the presence of an ANO vortex, a baryon emerges on the vortex at a $\mu_B$ lower than what is demanded by a conventional nucleon. 

If we are to apply our results to study the phase of a bulk of baryonic matter, effects of the magnetic field need to be further investigated. 
An Abrikosov vortex lattice (AVL) shall occupy a certain region of the phase diagram with one parameter being the magnetic field 
\cite{Gronli:2022cri}. 
A more detailed study on the role played by our proposed baryonic vortex in such a phase diagram requires further information of the vortex lattice, especially its transverse density which is essentially infinitesimal in our present one-soliton scenario. 

\section{Summary and Outlook}\label{sec:summary}

In this work, we have investigated ANO vortices in ChPT at leading order, incorporating the effects of pion mass $m_\pi$, isospin chemical potential $\mu_I$, and the coupling to Maxwell electrodynamics. 
The previous study~\cite{Qiu:2024zpg} analyzed such vortices in the chiral limit ($m_\pi = 0$), in which case the $\pi^0$
inside the $\pi^\pm$ vortex core varies linearly along the vortex string (the $z$-axis).
In contrast, we have shown that the inclusion of $m_\pi\neq0$ leads to 
distinct and richer structures. 
Particularly, with both charged and neutral pion condensates in the bulk, $\pi^0$ is no longer confined within the vortex core. 
As a result, the system accommodates two different kinds of vortices among the configuration: a local (ANO-like) vortex associated with the winding of $\pi^\pm$, and a global vortex corresponding to the winding of $\pi^0$, attached to a $\pi^0$ domain wall (a.k.a chiral soliton). 

We have demonstrated that when the baryon chemical potential $\mu_B$ exceeds a critical value $\mu_c$ that depends on the isospin chemical potential $\mu_I$, an ANO vortex along $z$-axis becomes topologically linked with a closed neutral pion string. We have shown that such a linked configuration carries a baryon number, thereby realizing the homotopy that is identical to a Skyrmion. 
This interpretation is supported also by previous efforts which establish that the linking number of such vortices corresponds to the topological baryon charge~\cite{Gudnason:2020luj,Gudnason:2020qkd}. 
In this way, our study provides a novel mechanism of Skyrmion formation without a Skyrme term. 
Such findings may open new
horizons for understanding baryonic structures in dense matter such as neutron stars and heavy-ion collisions, where isospin and baryon asymmetries coexist under extreme conditions.

In this paper, we have not yet taken into account an external magnetic field, but it will be quintessential for applications in physical
contexts of neturon stars and heavy-ion collisions.
When an applied external magnetic field reaches a critical value, an Abrikosov vortex lattice (AVL) 
of $\pi^\pm$ becomes the ground state, 
in systems with 
$\mu_I\neq 0$ and $\mu_B=0$, known from previous study~\cite{Adhikari:2015wva,Adhikari:2018fwm,Adhikari:2022cks,Gronli:2022cri}.
In particular, in ref.~\cite{Gronli:2022cri} 
a phase diagram in the plane of $\mu_I$ and applied magnetic field was 
worked out, separating  
the AVL and CSL, against the QCD vacuum. 
Our results in this paper imply that  
the AVL (with zero baryon number) transits to a lattice of linked baryonic vortices (carrying a finite baryon number) when $\mu_B$ exceeds the critical value as a function of $\mu_I$, as shown in fig.~\ref{fig:phase-diagram}. 
In this context, if the magnitude of the magnetic field is added as a third dimension,
fig.~\ref{fig:phase-diagram} is exactly the slice (of a certain 3D phase diagram) at the critical magnetic field of AVL, where the transition from QCD vacuum to nontrivial vortex configurations would begin with an AVL of well-separated vortices.
In some sense, a transition from the AVL further into the linked baryonic vortex lattice addressed here is comparable to a transition from the CSL to the domain-wall Skyrmion phase. 
The former roots in the $\mu_I$-induced $\pi^\pm$ condensate while the latter originates from the $\mu_BB$-dependent $\pi^0$ domain wall, both ending up into Skyrmionic configurations.
On the other hand, in a baryonic vortex phase, if we decrease $\mu_I$ to eventually less than $m_\pi$, we suppose that the $\pi^\pm$ condensate in the bulk would disappear and then the vortex-Skyrmion would transit to a conventional Skyrmion in the magnetic field, which may exhibit a ``pancake'' structure
\cite{Son:2007ny,Amari:2025twm} close to fig.~\ref{fig:3D-link} schematically. 
If we consider the limit of $\mu_I \to 0$, we conjecture the phase would eventually transit to the CSL 
for a sufficiently large $\mu_B B$.
How occur these phase transitions involving $\mu_I$, $\mu_B$, and the magnetic field in low energy hadronic matter is an intriguing issue to explore further.

Another noteworthy message from our results is, in dense hadronic matter with isospin asymmetry, such as 
neutron stars, for $\mu_B\gg\mu_I$ the baryonic vortex phase likely exists in reality. Importantly, the vortex accommodates a conserved magnetic flux, providing a possible microscopic scenario that explains the long-lived strong magnetic fields inside neutron stars.
Estimating the strength of such magnetic fields held by the baryonic vortices needs macroscopic theories capturing the equation of state, which shall be a promising outlook based on the present study.

\section*{Acknowledgments}

This work is supported in part by JSPS Grant-in-Aid for Scientific Research KAKENHI Grant No. JP22H01221 and JP23K22492 (M.~N. and Z.~Q.)
and by the Deutsche Forschungsgemeinschaft under Germany's Excellence Strategy - EXC 2121 Quantum Universe - 390833306.
The work of M.~N. is supported in part by the WPI program ``Sustainability with Knotted Chiral Meta Matter (WPI-SKCM$^2$)'' at Hiroshima University.

\bibliographystyle{jhep}
\bibliography{references}

\end{document}